\documentclass[12pt,preprint]{aastex}
\usepackage{graphicx}
\slugcomment{submitted to \apj Letters}

\shorttitle{Progenitors for SNe Ia  \& Binary evolution}
\shortauthors{Chen et al. }

\begin{document}
%\title{The influence of non-conservative merging and hydrodynamic simulation limits on the double-degenerate model for Type Ia supernovae}

\title{Some constraints on the  lower mass limit for  double-degenerate progenitors of Type Ia supernovae}

\author{X. Chen$^{1,2}$, C. S. Jeffery$^{3,4}$, X. Zhang$^{3}$ and Z. Han$^{1,2}$,}
\affil{$^1$National Astronomical Observatories / Yunnan
Observatory, Chinese Academy of Sciences, Kunming, 650011, China\\
       $^2$ Key Laboratory for the Structure and Evolution of Celestial
Objects, Chinese Academy of Sciences, Kunming, 650011, China\\
$^3$Armagh Observatory, College Hill, Armagh BT61 9DG\\
$^4$School of Physics, Trinity College Dublin, Dublin 2, Ireland }
\email{xuefeichen717@hotmail.com}

\begin{abstract}
Recent theoretical and observational studies both argue that the
merging of double carbon-oxygen white dwarfs (WDs) is responsible
for at least some Type Ia supernovae (SNe~Ia). Previous (standard)
studies of the anticipated SN birthrate from this channel have
assumed that the merger process is conservative and that the
primary criterion for explosion is that the merged mass exceeds
the Chandrasekhar mass. Han \& Webbink (1999) demonstrated  that
mass transfer and merger in close double WDs will in many cases be
non-conservative.  Pakmor et al. (2011) further suggested that
 the merger process should be
violent in order to initiate an explosion. We have therefore
investigated how the SN~Ia birthrate from the double-degenerate
(DD) channel is affected by these constraints. Using the
binary-star population-synthesis method, we have calculated the DD
SN~Ia birthrate under conservative and non-conservative
approximations, and including lower mass and mass-ratio limits
indicated by recent smoothed-particle-hydrodynamic calculations.
The predicted DD SN~Ia rate is significantly reduced by all of
these constraints. With dynamical mass loss alone (violent merger)
the birthrate is reduced to 56\% of the conservative rate.
Requiring that the mass ratio $q>2/3$ further reduces the
birthrate to 18\% that of the standard assumption. An upper limit
of 0.0061 SNuM, or a Galactic rate of $4.6 \times 10^{-4}{\rm
yr}^{-1}$, might be realistic.
\end{abstract}


\section{Introduction}
The double-degenerate (DD) merger was suggested as a possible
channel for type Ia supernovae (SNe~Ia) in the early 1980's
\citep{tutu81,it84,web84}. In this DD model, two carbon-oxygen
(CO) white dwarfs (WDs) can produce a SN~Ia while merging if their
total mass is larger than the Chandrasekhar  mass ($M_{\rm ch}$).
The DD model can naturally explain the lack of H and He emission
in the spectra of SNe~Ia and some super-luminous SNe~Ia, but it
has a major difficulty in explaining the similarities of most
SNe~Ia since the merger mass has a relatively wide range, $\sim
1.4-2.0M_\odot$ \citep{wang10}\footnote{It can be argued, however,
that in the DD scenario it is the primary (exploding) WD which
matters most, not the total mass, since it is the detonation in
the exploding WD which is primarily responsible for synthesizing
$^{56}{\rm Ni}$ \citep{pak10,pak12}.}. Meanwhile, although the
SNe~Ia rate predicted from the DD model is comparable to that of
observations \citep{yun94,han98,nel01,rui09,wang10,yu10},
theoretical studies suggest that the merger of two WDs is more
likely to lead to an accretion-induced collapse (AIC) to form a
neutron star \citep{nomo85,saio85, tim94}. Consequently, the DD
model has been disadvantaged in comparison to the
single-degenerate (SD) model for a long time \citep{wang12}.

\citet{pie03} suggested that, under the right conditions, the DD
merger process could be quite {\it violent} and lead to a SN Ia
explosion rather than AIC (see also \citet{yoon07,shen12}). Fully
three-dimensional simulations of a violent merger of two CO WDs by
\citet{pak10,pak11} show that the merger of two equal-mass CO WDs
($\sim 0.9M_\odot$) can explain the formation of sub-luminous
1991bg-like events (see also \citet{van10}). The \citet{pak12}
simulation of a DD merger  with masses of $1.1M_\odot$ and
$0.9M_\odot$  shows good agreement with properties of normal
SNe~Ia. Observationally, some known  or potential DD systems such
as WD 2020-425 \citep{nap07}, V458 Vulpeculae \citep{rod10},
SBS1150+599A \citep{tov10} and GD687  \citep{gei10} represent good
candidates for SN~Ia progenitors since they have  total masses
close to $M_{\rm ch}$ and orbital periods short enough to merge
within a Hubble time. Furthermore, extensive searches have found
no surviving companion to the Type Ia supernovae responsible for
SNR 0509-67.5 \citep{sch12} or SN 1572 \citep{ker12}, whence it is
argued that the progenitors must have been DD systems.   Hence, it
appears that at least some, if not most, SNe~Ia come from  DD
mergers.

Previous studies for the DD model generally assume that a SN~Ia is
produced if the total mass of a DD system is larger than  $M_{\rm
ch}$. This assumption implies that the DD merger process is always
conservative. However, if accretion is spherically symmetric, mass
transfer can be dramatically non-conservative both during stable
mass transfer and in a violent merger. \citet{han99} (HW) studied
the stability and energetics of mass transfer in double WDs. They
showed that the expelled mass fraction of an interacting double WD
depends on the component masses and can be as large as 50\% of the
donor for a system with masses of $1M_\odot$ and $0.5M_\odot$.
Meanwhile, smoothed-particle-hydrodynamic (SPH) simulations
\citep{yoon07,pak11} have shown that not all double CO WDs with a
total mass larger than $M_{\rm ch}$ can produce SNe~Ia. For
example, \citet{pak11} find that only double CO WDs in which the
more massive white dwarf exceeds $\sim 0.9M_\odot$ and the mass
ratio roughly exceeds 0.8 robustly reach the conditions required
for initiating a detonation, leading to a SN~Ia explosion.
Obviously, some additional constraints should be placed on the DD
model for triggering SNe~Ia\footnote{From the point of view of
preventing an off-center C ignition, \citet{yoon07} favored less
massive double CO WDs to produce SNe~Ia. Since an off-center C
ignition is almost inevitable \citep{shen12}, and much of  the
physics for the thermal evolution of merger remnants is unclear,
we only consider here the constraints from \citet{pak11}. }.

In sect.2 we investigate the merger mass of double WDs including
non-conservative mass transfer, and we examine the impact of these
constraints and those arising from SPH simulations on the SNe~Ia
birthrate in sect.3. Conclusions are drawn in sect.4.

\section{The Merger Mass}
As the orbit of  a double WD system decays due to the emission of
gravitational radiation/waves, the less massive component will be
the first to fill its Roche lobe and become the mass donor; the
more massive component becomes the accretor. Unlike non-degenerate
donor stars in other compact binaries, the WD donors in double WDs
lie deeper within the potential well of the accreting star,
resulting in important consequences for mass transfer. Assuming
energy conservation and hence no energy sources other than
gravitational potential energy are at play, HW analyzed the
initial stability of this mass transfer and the fraction of the
mass-transfer stream to be expelled from the binary. From this
point of view, the mass transfer is conservative if the accretion
luminosity is less than the Eddington limit. Any accretion
luminosity in excess of the Eddington limit is assumed to be
absorbed in the accretion flow. In terms of energetics, radiative
losses represent accretion energy not used to power a mass outflow
 (HW: Eq (21)). The fraction of matter accreted by the
primary is used to power the outflow and can be determined from
HW: Eq (22). For component masses $M_1$ (accretor) and $M_2$
(donor), we adopt HW: Eqs (17), (22) and (23) to calculate the
mass transfer rate, $\dot M_{\rm 2}$, and the fraction of matter
to be accreted by the primary\footnote{Note that the model of HW
takes the accretion to be spherical. If mass transfer is through a
disk, then excess energy can be radiated away from the poles and
$\beta$ may not be appropriate.}, $\beta$.

There are two regimes for computing the final total mass of the
merger, $M$. If the merger process is dynamical,  $\beta$ can be
applied as a single step process and $M = M_1 + \beta M_2$. If
mass transfer is stable, $M$ is obtained by integrating over the
mass-transfer epoch\footnote{The fitting formulae and a table for
the merger mass from the two regimes for various component masses
may be obtained by email from xuefeichen717@hotmail.com.}, using
$M = M_1 + \int \beta(t){\dot M_2(t)} {\rm d}t$.

\section{Type Ia supernovae from the DD model}

\subsection{The formation of double CO WD systems}

To investigate the SN~Ia birthrate due to DD mergers, we require a population of
 close double CO WD binaries, to be formed from interactions in a population of
primordial binary stars (e.g. \citet{han98}). We first perform a
Monte-Carlo simulation to obtain a stellar population, then evolve
them in a rapid binary evolution code (RBEC) \citep{hur00,hur02}
to obtain a sample of close CO WD systems.

In the Monte Carlo simulation, all stars are assumed to be members
of binaries and have circular orbits. The primaries follow the
initial mass function of \citet{mil79} and are generated according
to the formula of \cite{egg89}, in the mass range 0.08 to
$100M_\odot$. The secondary mass, also with a lower limit of
$0.08M_\odot$, is then obtained from a constant mass-ratio
distribution. The distribution of orbital separations $a$ is taken
to be constant in ${\rm log} a$ for wide binaries. This separation
distribution has been used in many Monte-Carlo simulations and
implies an equal number of wide binary systems per logarithmic
interval and approximately 50 per cent of stellar systems with
orbital periods less than 100 yr \citep{han98}.
Long-orbital-period binaries are effectively single stars.

The formation of close double CO WD systems depends significantly
on the critical mass ratio for dynamical instability $q_{\rm c}$
and common envelope (CE) evolution \citep{han98}. A binary
experiences stable Roche lobe overflow (RLOF) when $q\leq q_{\rm
c}$ and CE evolution otherwise, where $q=$ donor/accretor. In this
study we adopt equation (57) of \citet{hur02} for $q_{\rm c}$ when
the mass donor is on the first or asymptotic giant branch (AGB),
and let $q_{\rm c}=4$ when the mass donor is on the main sequence
or in the Hertzsprung gap, as supported by detailed binary
evolution studies \citep{han00,chen02,chen03}. For CE evolution,
we use the standard energy formalism \citep{heu76,web84,livio88},
that is, the CE is ejected if $\alpha_{\rm ce} \Delta E_{\rm
orb}\ge E_{\rm bind}$, where $\Delta E_{\rm orb}$ is the orbital
energy released, $\alpha_{\rm ce}$ is the CE ejection efficiency,
and $E_{\rm bind}$ is the binding energy of the envelope and  can
be written as $GM_{\rm d}M_{\rm env}/(\lambda R_{\rm d})$ (where
$M_{\rm d}$, $M_{\rm env}$ and $R_{\rm d}$ are the mass, envelope
mass and radius of the donor, respectively, and $\lambda$ is a
structure parameter that depends on the evolutionary stage of the
mass donor). We combine $\alpha_{\rm ce}$ and $\lambda$ into one
free parameter, setting $\alpha_{\rm ce}\lambda= 1.5$ to reproduce
the number of the DD objects in the Galaxy as in previous studies
\citep{wang09b}.

RBEC distinguishes three types of WDs, namely helium (expected
only in binaries), CO and oxygen-neon (ONe), respectively. The
critical condition for CO and ONe WDs is the core mass at the base
of AGB, $M_{\rm c,BAGB}$. CO WDs are produced when $M_{\rm
c,BAGB}<1.6M_\odot$, and ONe WDs when $1.6M_\odot \le M_{\rm
c,BAGB}\le 2.5M_\odot$, where $M_{\rm c,BAGB}$ is given in Eq (66)
of \citet{pol98}. The mass of the CO core, and the final mass of
CO WDs is determined by the $L-M_{\rm c}$ relation, i.e. Eqs(37)
and (39) of that paper.

\subsection{Constraints on the progenitors of SNe~Ia}

There are two basic constraints on double CO WDs as the
progenitors of SNe~Ia in the conservative case: (i) $M_1+M_2 >
M_{\rm ch}$\footnote{We adopt $M_{\rm ch} = 1.378M_\odot$
following \citet{han04,meng09,wang09a}.}, and (ii) the CO WDs are
close enough to merge in a Hubble time. The timescale for two
components to merge by gravitational wave radiation, $t_{\rm GW}$,
is written as \citep{lan71}
\begin{equation}
t_{\rm GW}=8\times10^7\times {(M_1+M_2)^{1/3}\over M_1M_2}P^{8/3},
\end{equation}
where $P$ is the orbital period in hours, $t_{\rm GW}$ in years
and $M_1$, $M_2$ in solar mass.

In the {\it non}-conservative case, (i) should be redefined as the
final mass $M > M_{\rm ch}$ {\bf (see sect. 2)}. Furthermore, AIC
should be avoided in order to successfully trigger a SN Ia. We
therefore need additional constraints from dynamical simulations.
\citet{pak10,pak11,pak12} showed that an initial detonation at the
onset of C ignition can lead the DD merger to avoid AIC and
explode, and that DDs with $M_1\geq 0.9M_\odot$ and a mass ratio
$M_1/M_2 \gtrsim 0.8$ can achieve conditions for initiating such a
detonation. Their studies also showed that mergers with $M_1
\simeq 0.9M_\odot$ are very promising candidates for explaining
{\it subluminous} SNe~Ia, while a {\it normal} SN~Ia can be
produced from the double CO WDs with $M_1=1.1M_\odot$. Since the
SN Ia luminosity is determined by the amount of $^{56}{\rm Ni}$
synthesized in the explosion, these simulation results seem to
indicate that the primary mass is related to the final production
of $^{56}{\rm Ni}$, and therefore determines whether a normal or
sub-luminous SN~Ia is produced. Meanwhile, we {\it do} see a lot
of variation in SN Ia lightcurves, and thus one might expect to
see a variety among exploding WD masses provided by the DD models.
In the absence of other constraints, we study the SN~Ia birthrate
assuming two lower mass limits for the DD accretor, i.e. $M_1>
0.9$ or $1.0M_\odot$, respectively.

Another constraint from the SPH simulations is the mass ratio of
the CO DDs, $q=M_2/M_1$. It has a critical value, $\approx 0.8$,
below which the merger is not violent enough to ignite a
detonation (Pakmor et al. 2011)\footnote{Pakmor et al. (2011)
found that the critical value of $q$ probably changes with primary
WD mass; i.e. more massive primaries can merge with lower mass
ratios. }. The relation between the violence of a merger and the
mass ratio can be simply understood from Fig. 2 of HW. In the
dynamical instability region, $\zeta_{\rm ad}-\zeta_{\rm L}<0$,
where $\zeta_{\rm ad}$ and $\zeta_{\rm L}$ are the adiabatic
mass-radius exponent of the donor and the mass-radius exponent of
its Roche radius, respectively. With the mass ratio closer to
unity, the value of $\mid\zeta_{\rm ad}-\zeta_{\rm L}\mid$ becomes
larger and larger. Since it is a negative number, the process of
self-amplifying $\dot M_2$ (Eq (1) in that paper) is thus faster
and faster, resulting in a more and more violent merger process.
We arbitrarily relax the critical mass ratio from 0.8 to 2/3 in
our study to give an upper limit for the SN~Ia birthrate from the
SPH simulations.

In brief, based on the sample of double CO WDs which can merge in a
Hubble time, we compute the SN~Ia birthrate for the following cases :

(i) $M=M_1+\beta M_2\ge M_{\rm ch}$ ;

(ii) $M =M_1+\int \beta(t)\dot M_2(t) {\rm d}t\ge M_{\rm ch}$;

(iii) As case (ii) with $M_1\ge0.9M_\odot$. We choose case (ii) as
the basic constraint  since it includes a larger parameter space
than case(i) (see Fig.1). The overall SNe Ia rate from case (ii)
can then be considered as an upper limit from the non-conservative
assumption.

(iv) similar to (iii) but $M_1\ge1.0M_\odot$, and

(v) an additional constraint, $q=M_2/M_1\geq 2/3$, is placed on
case (iii).

Figure 1 shows these constraints in the $M_1-M_2$ plane (left
panel), as well as CO DDs with a total mass $M_1+M_2
> M_{\rm ch}$ which can merge in a Hubble time (right panel).
To obtain these DDs, we generate a population of $10^7$ binaries
(with a total mass of $\simeq 1.26\times 10^7M_\odot$) as
described in sect.3.1 and evolve them in RBEC. These DDs cover a
wide range in the $M_1-M_2$ plane, and also a large range of
$\beta$. For $q=2/3$, the $\beta$ decreases from 0.7 to 0.54 as
$M_1$ increases from 0.827 to $1.2M_\odot$. The overall parameter
spaces of SN~Ia progenitors in cases (i)-(iv), i.e.  to the right
of the red lines, are reduced by different degrees relative to the
conservative assumption, indicating corresponding decreases of the
DD SN~Ia rate under the non-conservative assumption and with
dynamical constraints.

\begin{figure}
\includegraphics[width=5.5cm,angle=270]{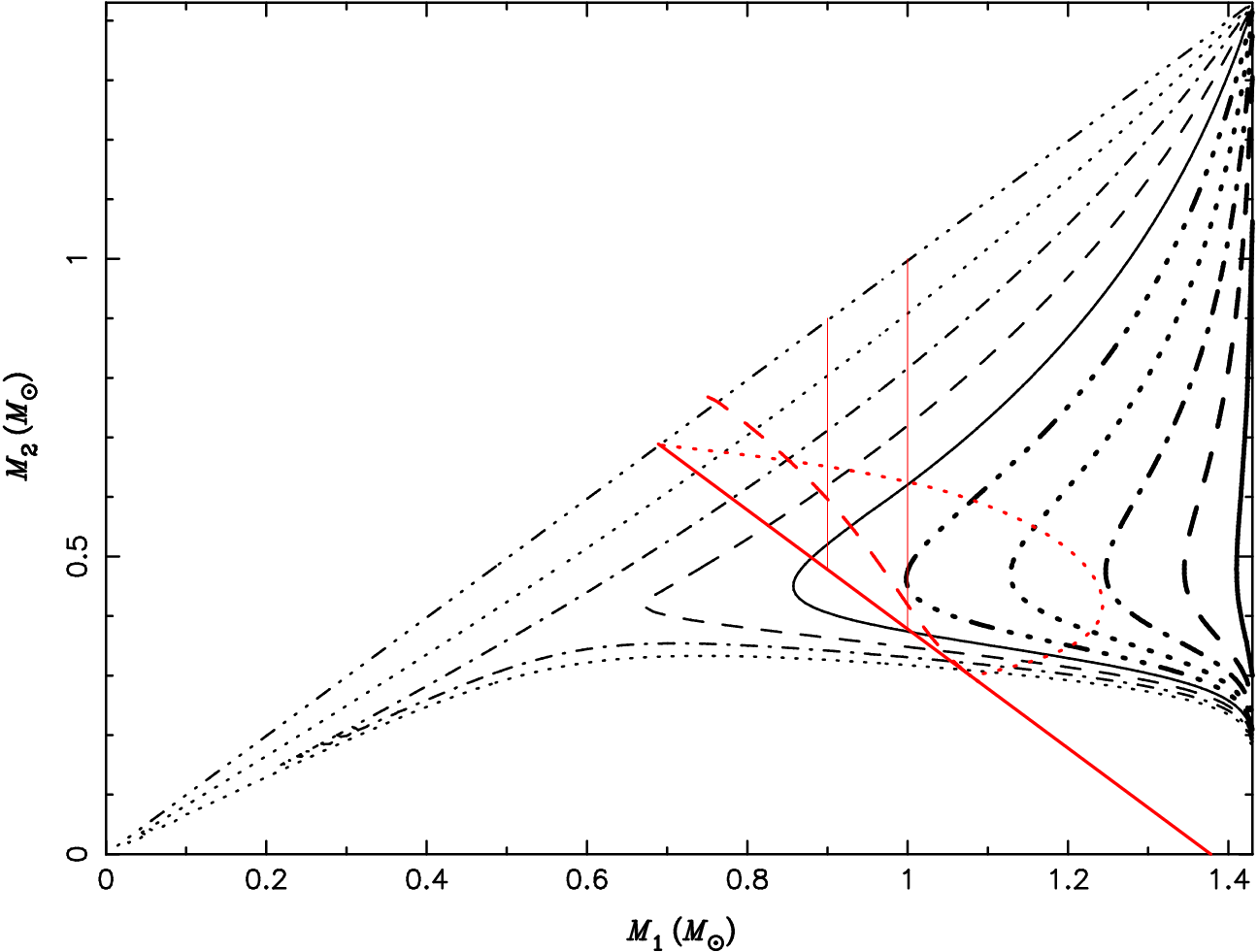}
\includegraphics[width=5.5cm,angle=270]{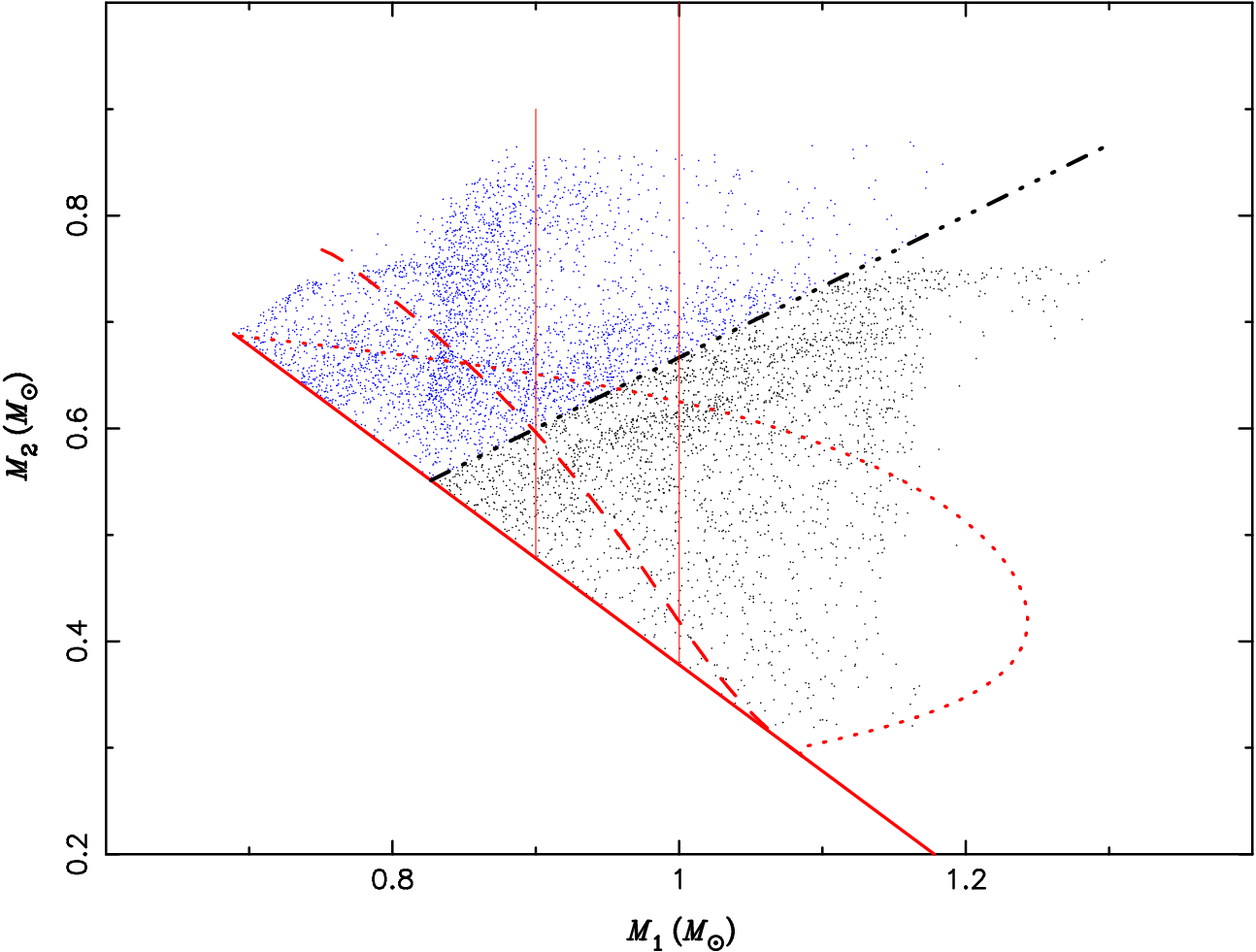}
\caption{Left panel: constraints on the progenitors of SNe~Ia
of double CO WDs in the $M_1-M_2$ plane. The black lines are
contours of the ejected mass fraction $1-\beta$ with values of,
from left to right, 0.0, 0.1, ... and 0.9, respectively. The red
lines mark contours of merger mass $M_1+M_2=M_{\rm ch}$
(conservative case, solid line), $M_1+\beta M_2=M_{\rm ch}$
(non-conservative, dotted line) and $M_1 + \int \beta(t) \dot
M_2(t) dt=M_{\rm ch}$ (non-conservative, dashed line). The
vertical lines give the boundaries for $M_1=0.9$ and $1.0M_\odot$.
The areas to the right of each colour line represent $M\geq M_{\rm
ch}$ and hence DD SN~Ia progenitor candidates under the
corresponding constraints. Right panel: CO DDs from our
population synthesis with $M_1+M_2>M_{\rm ch}$ which can merge
within a Hubble time, the blue dots have mass ratios
$q=M_2/M_1\geq 2/3$. The black dash-dot-dot-dotted line shows the
boundary of $q=2/3$, and the red lines are the same as those in
the left-hand panel.
 \label{sample}}
\end{figure}

\subsection{Birth rates of SNe~Ia}

The CO WDs above are used to study the SN~Ia birthrates in various
cases, which are finally normalized to (a) a single starburst of
$10^{11}M_\odot$  and (b) a constant star formation rate of
$5M_\odot {\rm yr}^{-1}$ over the past 15~Gyr, to resemble our
Galaxy. The results are shown in Fig.2. We see that, in the
conservative case, the predicted SN~Ia birthrate is 0.033
SNuM\footnote{1 SNuM$ =1 {\rm SN} (100{\rm yr})^{-1}(10^{10}{\rm
M_\odot})^{-1}$} (corresponding to a Galactic birth rate of
$2.5\times 10^{-3}{\rm yr^{-1}}$) for the constant star-formation
at 15 Gyr (left panel). In the non-conservative case, the
predicted overall rate is reduced to 0.019 SNuM ($1.4\times
10^{-3}{\rm yr^{-1}}$) and 0.024 SNuM ($1.8\times 10^{-3}{\rm
yr^{-1}}$) for cases (i) and (ii), respectively. If we introduce
additional restrictions from dynamical simulations, the overall
SN~Ia rate is further reduced to 0.017 SNuM ($1.3 \times 10^{-3}
{\rm yr}^{-1}$), 0.0099 SNuM ($7.4 \times 10^{-4} {\rm yr}^{-1}$)
and 0.0061 SNuM ($4.6 \times 10^{-4} {\rm yr}^{-1}$) for cases
(iii)-(v), respectively. The birthrate in case (v) is only about
18\% of that of the conservative case. Note that the SN Ia
birthrate from case (v) might/could represent an upper limit based
on constraints from recent SPH simulations (see sect.3.2).

Cases (i)-(iv) present similar delay-time distributions to that of
the conservative case, that is, a peak around $0.8-3.2 \times 10^8
{\rm yr}$ with a tail decaying as  $t^{-1}$, where $t$ is the
delay time (Fig.~2, right). The shape of this distribution is
consistent with both theoretical analyses and observations (see
Maoz et al. 2011; Wang \& Han 2012 and references therein).
However, in case (v), one sees a small peak around ${\rm log}
t({\rm yr})=8.7$ and a `dip' just before ${\rm log} t({\rm
yr})=8.5$.  RBEC produces CO WDs nearly uniformly distributed in
orbital period. The CE+CE channel produces extremely short orbital
periods and the RLOF+CE channel gives relatively longer orbital
periods, with some overlap between. Requiring $q \ge 2/3$ removes
low-$M_2$ binaries, mainly from the RLOF+CE channel. These
low-$M_2$ binaries are produced in two ways; i) the AGB
progenitors of $M_2$ before the CE phase have low masses and
generally lower binding energies and produce double CO WDs with
longer orbital periods for a given $M_1$, ii) binaries with short
orbital periods before CE ejection produce double CO WDs with
short orbital periods. Hence the $q>2/3$ constraint removes both
the longest- and shortest-period binaries from the RLOF+CE
channel, leaving a relatively larger number of progenitors with
orbital periods of $\sim 3$ hrs upon emerging from the CE+CE
channel (corresponding delay times of ${\rm log} t=8.7$). The
`dip' arises between the surviving CE+CE and RLOF+CE samples since
short-orbital-period systems from the RLOF+CE channel have been
removed.

\begin{figure}
\includegraphics[width=5.5cm,angle=270]{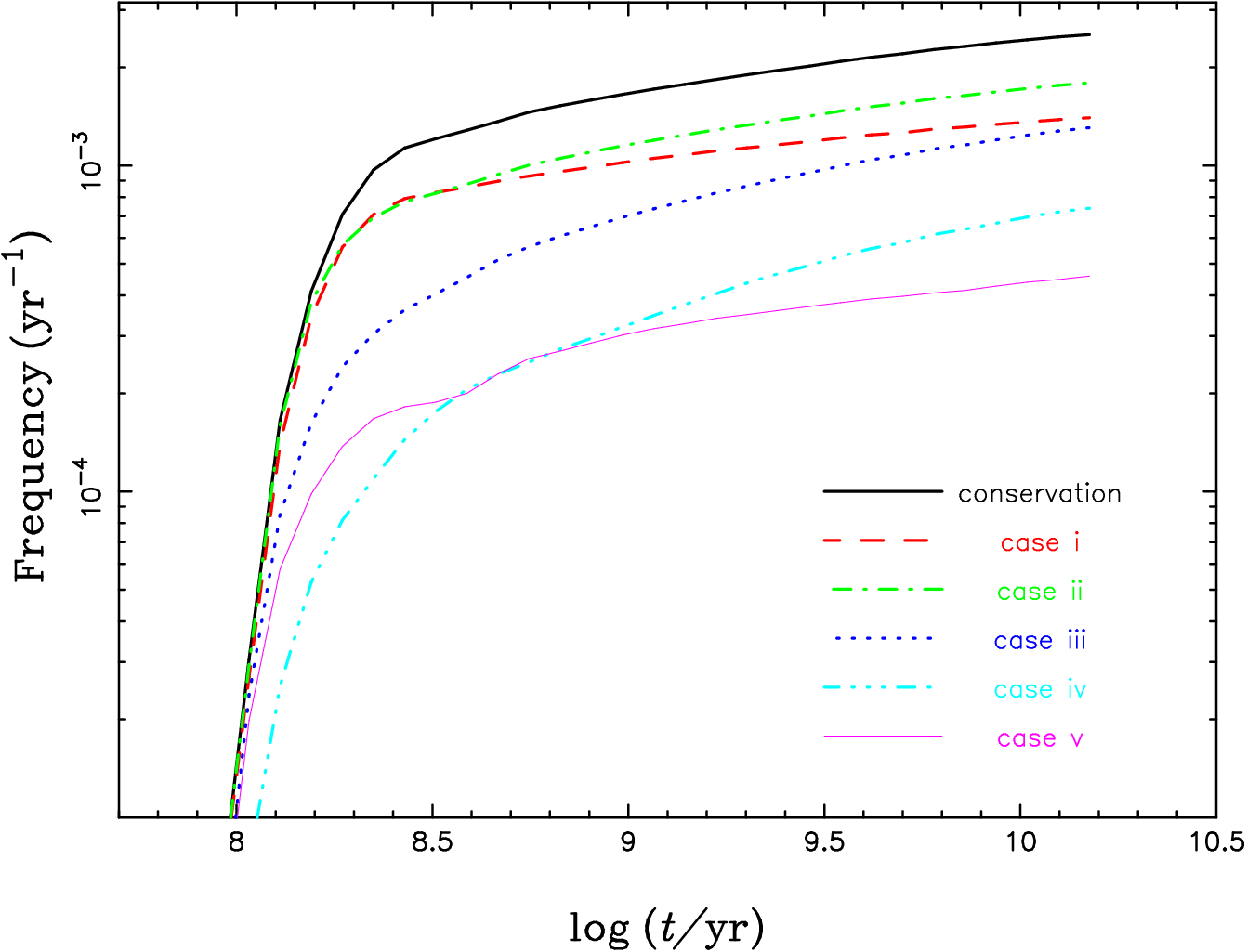}
\includegraphics[width=5.5cm,angle=270]{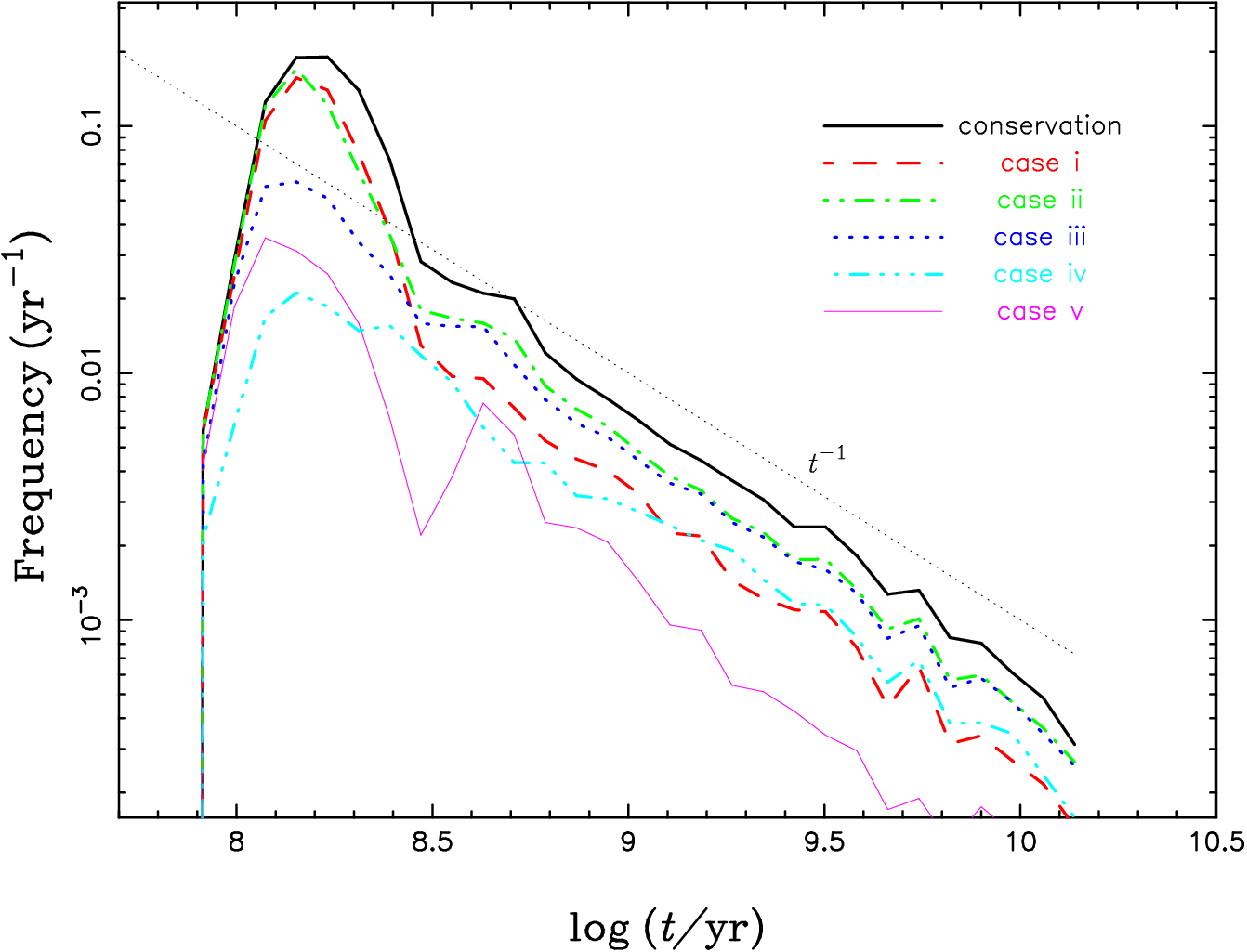}
\caption{Left panel: the evolution of birthrate of SNe~Ia for
a constant Pop I star formation rate ($5M_\odot {\rm yr}^{-1}$)
from the DD model. Right panel: as the left
panel for a single starburst of $10^{11}M_\odot$.
 \label{rate}}
\end{figure}

\section{Conclusion}
We have shown that the DD model for SNe~Ia is less effective than
previously assumed if we consider the non-conservative nature of
the DD merger process and constraints on triggering SN~Ia
explosions provided by recent SPH simulations. The overall SN~Ia
rate is reduced to a value $\approx 0.56 - 0.72 $ that of previous
studies (or $0.018-0.024$ SNuM) by introducing the
non-conservative approximation, while it is likely below 0.18 of
that value (or $<0.006$ SNuM) if we also consider the SPH
constraints.

Theoretical estimates of the SN~Ia rate from the SD model show a
wide range of values from $\approx 0.001$ SNuM \citep{rui09} to
$>0.01$ SNuM ( Han \& Podsiadlowski 2004) (see also Mennekens et
al. 2010), which strongly depends on the model assumptions and
input parameters used in various studies. \citet{wang10} performed
a study using the same synthesis model and Galactic (constant)
star formation rate as used in this study, and found a rate of
0.029 SNuM ($2.15\times 10^{-3}{\rm yr^{-1}}$). By direct
comparison, the DD model appears unlikely to be better than the SD
model in explaining the SN~Ia birthrate.

It was found by Pakmor et al. (2012) that an extended envelope
enshrouding the merging WDs is not produced in violent merging
events. However, we consider the possibility that a hot extended
envelope may arise following a non-violent non-conservative DD
merger. Since the disrupted material of the secondary must also
contain most of the original orbital angular momentum, the
outermost layers of the merger probably form a centrifugally
supported disc. Our study therefore suggests that the lost or
unaccreted material from the DD merger would take the form of a
hot envelope plus disk, similar to that indicated by the SPH
simulations of \citet{yoon07} in which the scale of the hot
envelope is small (i.e.  $\le 10^{10} {\rm cm}$). The absence of
early ultraviolet-optical emission in SN 2011fe \citep{nug11} is
compatible with the DD model with a small extended envelope, while
a disk would cause some continuum polarization such as that
detected at red wavelengths in SN 2011fe \citep{smith11}, although
other  causes of polarization cannot be ruled out.

\acknowledgments  We thank the referee for his/her suggestions.
This work is partly supported by the NSFC (Nos. 10973036,
11173055, 11033008 and 11003003), the CAS (No. KJCX2-YW-T24 and
the Talent Project of Western Light ), and the Project of Science
and Technology from the Ministry of Education (211102).

\end{document}